# Community structure of copper supply networks in the prehistoric Balkans: An independent evaluation of the archaeological record from the 7th to the 4th millennium BC

M. Radivojević[*]

[1]McDonald Institute for Archaeological Research, University of Cambridge, Downing Street, CB2 3ER Cambridge, UK
[*]Corresponding author: mr664@cam.ac.uk

J. Grujić[*]

[2]Vrije University Brussels, Artificial Intelligence Lab, Pleinlaan 2, Building G, 10th floor & MLG, Département d'Informatique, Université libre de Bruxelles, 1050 Brussels, Belgium
[*]Corresponding author: jgrujic@vub.ac.be

Complex networks analyses of many physical, biological and social phenomena show remarkable structural regularities [1-3], yet, their application in studying human past interaction remains underdeveloped. Here, we present an innovative method for identifying community structures in the archaeological record that allow for independent evaluation of the copper using societies in the Balkans, from c. 6200 to c. 3200 BC. We achieve this by exploring modularity of networked systems of these societies across an estimated 3000 years. We employ chemical data of copper-based objects from 79 archaeological sites as the independent variable for detecting most densely interconnected sets of nodes with a modularity maximization method [4]. Our results reveal three dominant modular structures across the entire period, which exhibit strong spatial and temporal significance. We interpret patterns of copper supply among prehistoric societies as reflective of social relations, which emerge as equally important as physical proximity. Although designed on a variable isolated from any archaeological and spatiotemporal information, our method provides archaeologically and spatiotemporally meaningful results. It produces models of human interaction and cooperation that can be evaluated independently of established archaeological systematics, and can find wide application on any quantitative data from archaeological and historical record.

## 1. Introduction

Complex networks analyses have found a wide variety of applications in many disciplines [5], from economy [6] to neuroscience [7], and include explorations of the nature of diverse technological, biological and social systems [8-12]. Social networks are of particular interest since they are important for the development of cooperative interactions amongst humans [13]. Examples from studies of hunter-gatherer cooperation [14] reveal how social proximity shaped cooperative behaviour in addition to genetic and geographic relations. Moreover, homophily in cooperation and selective formation of network ties amongst humans emphasises that socially connected individuals tend to resemble one another in preferences, values, and beliefs [15].

In archaeological systematics, the depositions of similar associations of material, dwelling and subsistence forms distributed across distinct space-time are globally referred to as 'archaeological cultures'. Thus, archaeologists have mostly strived to either group or split distinctive archaeological cultures based on specific expressions of similarity and its reproduction across the defined time and space [16]. The major point of debate on this matter is that specific expressions of similarity or differences are to a great extent based on subjective estimates, initially outlined in the early 20$^{th}$ century, and commonly in relation to one material, pottery. As a result, what they represent, and at what resolution remains a major problem in the field of archaeology [17].

Here, we present a method that independently evaluates community structures in the Balkans between c. 6200 BC to c. 3200 BC as proxied by the exchange of copper objects smelted from local ores [18-20]. This segment of the archaeological record possesses the world's earliest known evidence for the evolution of extractive metal production [21, 22] and inferences about community structure help illuminate the existence of networks of metal production and exchange. Also, they provide the closest approximation of social and economic ties at the time.

Our analyses include methods from complex networks, and in particular modularity research on a comprehensive archaeological database of copper artefacts circulating the prehistoric Balkans in the first 3000 years since the emergence of copper mineral and metal use in Europe. The database we compiled includes 410 copper-based artefacts distributed across c. 270,000 km$^2$ in the northern Balkan Peninsula (Table S1). All artefacts come from 79 sites in total, of which 14 are multi-occupational, meaning that they existed through several chronological periods. Therefore, we have 93 site-periods in total, which correspond with the number of nodes in our networks (Table S2).

The application of networks analysis in archaeology has thus far been dominated by a few popular perspectives [23]. The focus of scholarship in this field has usually been on mutual properties of networked systems: scale-free networks, small world phenomena, power-law degree distribution and agent-based modelling, with emphasis set on identifying the key sites in a certain period [24-30]. Most studies derive from the application of geographic information systems (GIS), where major outcomes are geographical networks indicating communication routes between the studied sites, and a few involve the establishment of links between nodes of interest based on typological similarities in material culture (obsidian or ceramic) [31, 32] and their geochemistry [33]. Mills et al. [28] made a significant impact with their use of networks analyses in examining the interaction of ceramic, obsidian and spatial networks in the $13^{th}$ – $15^{th}$ century AD US Southwest, which set out the methodological framework for challenging traditional artefact attribution approaches to understanding the interrelationship of spatial, social and material variables.

Community structure (modularity) research, however, has not yet found its application in studying past social networks from archaeological data. This property stands for organisation of nodes in modules, where many links join nodes of the same module and comparatively few links join nodes with other modules or other parts of a network [3, 34]. Methods for modularity maximization have developed significantly, with success demonstrated in detecting modules in citation networks, food webs, and pollination systems [35, 36], amongst others. Being able to infer social groups in archaeological data using this network property would be a major step towards identifying patterns of human interaction in our past. We operate here under the premise that the supply networks of copper-based artefacts can reveal information relevant for the specific histories of their users, and hence reflect human behaviour [37]. Noteworthy is that metal objects, unlike obsidian for instance, are more challenging to trace back to their source, due to the various processes involved in metal extraction or (re)melting, which may fractionate the original chemical signature. Modern research addresses the origins of copper artefacts by measuring both lead isotope and trace element content, and works best by excluding potential ore deposits, rather than narrowing it down to a particular source [38]. Hence, studying human interaction though the circulation of metal objects carries more challenge than materials with less complex properties, which is why we opted for modularity research, an approach used here to explore the structure behind the network of copper objects with shared chemical signatures (Supplementary Information).

**Copper supply networks – previous research.** In their study on the provenance of prehistoric Balkan metallurgy, Pernicka et al. [18, 19] produced high-precision provenance data for more than 300 copper artefacts from this region. For the purpose of their research, they used average-link cluster analyses to group these artefacts into nine compositional groups-clusters, of which two clusters showed good agreement with the trace element pattern of two ancient copper mines in the Balkans: Majdanpek in eastern Serbia and Ai Bunar in central Bulgaria, both exploited throughout the $5^{th}$ millennium BC (see SI for details). The broad patterns of copper circulation at this time identified by Pernicka et al. [18, 19] include: a) strong regionalisation of copper production and trade patterns, and b) noticeable shifts in copper-making industry through time. In practice, the former was drawn from cases where prehistoric Serbian metal mostly came from eastern Serbian copper mines (e.g. Majdanpek), while the latter refers to the simultaneous collapse of one copper exploitation system (e.g.

central and eastern Bulgaria around 4100 BC) and the rise of another (e.g. eastern Serbia after c. 4100 BC). These studies provided insight into the copper exploitation and acquisition practices for the first c. 3000 years of copper use in this part of the world; however, little is known how and if these data reflect social networks of interaction, and can we use them to independently evaluate archaeological interpretations of social or economic structures that supported the production, use and deposition of copper objects within the observed time.

## 2. Methods

**Data.** We used published and unpublished sets of compositional data for 410 copper-based objects under consideration, spanning c. 6200 to c. 3200 BC (Table S1, dataset on https://doi.org/10.17863/CAM.9599). The time span between c. 6200 and c. 3200 BC was divided into seven periods, based on both absolute and relative chronology: Early/Middle Neolithic (Period 1, 6200-5500 BC), Late Neolithic (Period 2, 5500-5000 BC), Early Chalcolithic (Period 3, 5000-4600 BC), Middle Chalcolithic (Period 4, 4600-4450 BC), Late Chalcolithic (Period 5, 4450-4100 BC), Final Chalcolithic (Period 6, 4100-3700 BC) and Proto Bronze Age (Period 7, 3700-3200 BC) (Table S3). The data were assembled from publications of Pernicka et al. [18, 19], Radivojević et al. [20], Radivojević [39] and UK's AHRC-funded "Rise of Metallurgy in Eurasia" project (No. AH/J001406/1 hosted by the UCL Institute of Archaeology) [40] (see Supplementary Information).

**Community property (modularity) analysis**. In the course of our research, we designed two distinctive networks: one, that had *artefacts* for nodes (*Artefacts Network*), and the other, where archaeological *sites* acted as nodes (*Sites Network*). Our Artefacts and Sites networks were defined exclusively on data (selected trace elements for 410 copper artefacts) isolated from any geographical, cultural or chronological information, in order to secure an independent estimate of economic and social ties amongst copper-using societies in the Balkans within the observed time. Our network was built in two discrete steps: 1) we grouped the data in ten distinctive chemical clusters (Artefacts Network); 2) placed a connector between the sites that contain pairs of artefacts from the same cluster and analysed the modularity of the final network (Sites Network). In both steps we used the Louvain algorithm [4] to obtain community structures (modules), and bootstrapping to test the significance of acquired results. Essentially, we formed a bipartite network, where one type of nodes are artefacts and the other type of nodes are sites; this network is simple in design, since one artefact can belong to one site only. The information we use is transformed from the Artefacts Network to the Sites Network: the chemical composition of artefacts is employed for their clustering, and subsequently these clusters are used to link the sites where artefacts were discovered. Hence, the links in the Artefacts Network connect artefacts with a similar chemical signature, while the links in the Sites Network connect sites that yielded artefacts that belong to the same chemical cluster. Similar approach has been attempted before in a different field and type of dataset, for example for designing a network of authors and their publications [41].

*Artefacts Network – clustering the copper objects by trace element chemistry.* Each artefact in our study has a unique chemical composition, which besides predominant copper contains trace elements (usually below 1 wt% or 10,000 ppm). Of all trace elements, seven (As, Sb, Co, Ni, Ag, Au, Se) are considered as the indicators of the origin, or the chemical signature of copper ore (s)melted to make this artefact [38, 42]. This is due to the affinity of these elements to transit into molten copper metal during its extraction from copper ores [43], and such behaviour has already proved useful in tracing the origins of the 5$^{th}$ millennium BC copper metal artefacts from the Balkans analysed by Pernicka et al. [18, 19]. Their dataset makes a large part of our study collection (335 artefacts out of 410 in Table S1).

Theoretically, the goal of chemical clustering is to detect groups of copper artefacts whose compositional signature (here a string of 7 trace elements) is more similar within a group (or a cluster) than with compositional signature of copper artefacts - members of other groups (or clusters). In other words, the links that join copper artefacts of the same chemical cluster are based on compositional similarity, and they are comparatively stronger within a cluster of chemically similar artefacts than the links connecting these artefacts to other clusters. Since this compares closely to the definition of network modularity [3, 34], we designed the cluster analyses based on the principles of community structure research in networks. As mentioned above, the nodes in our Artefacts Network are copper artefacts, while links are defined using Euclidean distance of the vectors of trace elements (see below and Supplementary Information for more details).

The calculation of Euclidean distance with the original trace element values was met with two problems: a) they showed lognormal, instead of Gaussian distribution in our case (Fig. 1) and b) they were correlated to begin with (Fig. 2a). Starting with the former, the lognormal distribution of our data indicated that small values are predominant (Fig. 1), and computing distances between the original data would lead to losing information on variation in smaller values. Hence, in order to account for relative differences, we transformed the original values into logarithms. The logarithms of original data brought out clearly the correlations between chemical elements, like Sb and As, Au, Ag and Se, or Sb with Ag/Au/Se (Fig. 2b). The nature of compositional data is specific, since all variables have a sum of 1 or 100% (this is known as constant-sum constraint, see below). It means that individual variables in the compositional data do not vary independently – i.e. if one variable decreases, the proportion of the remaining must increase. Such an induced correlation may easily hinder the true relationships among variables (in our case trace elements), which is why the next step in our data processing was to eliminate these correlations. For this, we ran principal component analysis (PCA), a statistical procedure used to reduce the dimensionality of a dataset consisting of a large number of interrelated variables, while retaining the variation present in the dataset [44]. It is otherwise the same procedure as eigenvalues decompositions from linear algebra. The PCA removed these correlations (Fig. S1), preparing the output, now calculated as principal component scores (Table S1), for network analysis.

The straight approach to PCA with original compositional data has already been known as fraught with difficulties because of the marked curvature often displayed by such datasets and the constant-sum constraint that each compositional vector must satisfy [45, 46]. Aitchison [45] proposed a way around these constraints by arguing that the best way to compute principal components out of restricted types of data (e.g. in allometry, or compositional data) is to use logarithms of the original data. This supports the treatment of our original data, although in our case it was also evident as a necessity from lognormal distribution (Fig. 1). A disadvantage of his approach was in that it could not handle zeros (0), which in our case was about to lead to losing a small handful of objects where particular trace elements were not detected (or were below the detection limit of the analytical instrument). An alternative, however, was to replace zero values with a small positive number, which is what we did before transforming the original values into logarithms. Our small positive number was smaller than the detection limit of any of the analysed elements (0.0001).

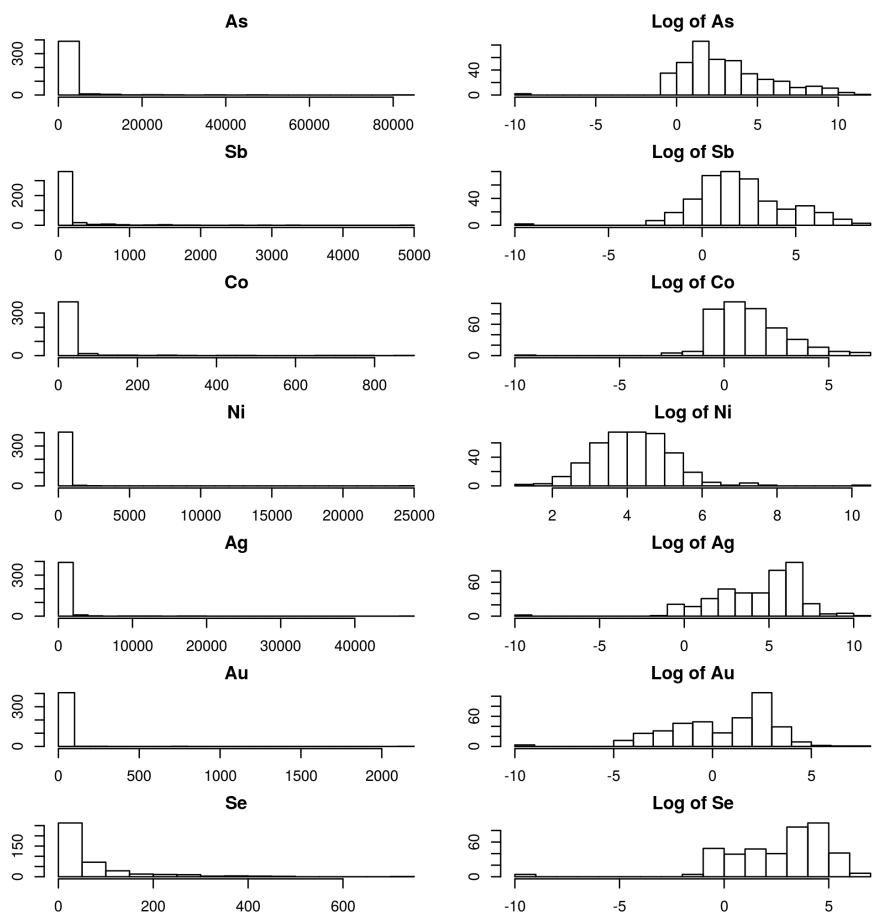

FIG. 1. Distribution of trace elements before (left) and after (right) the logarithmic transformation.

In the following step, the principal component scores (Table S1) were used to calculate the Euclidean distance between all pairs of artefacts. For this, we followed the rationale below: if $\vec{a}$ is a principal component vector of one artefact and $\vec{b}$ is a principal component vector of another artefact, the distance between the two artefacts will be defined as Euclidean distance between these two vectors as:

$$d(\vec{a}, \vec{b}) = \sqrt{\sum_{i=1}^{n}(a_i - b_i)^2}$$

Hence, the links in our Artefacts Network were defined as $1/d^2$ ($d$ = Euclidean distance). We acquired the number of clusters with Louvain algorithm [4], which is a high-performance method in complex networks analyses for identification of community structures.

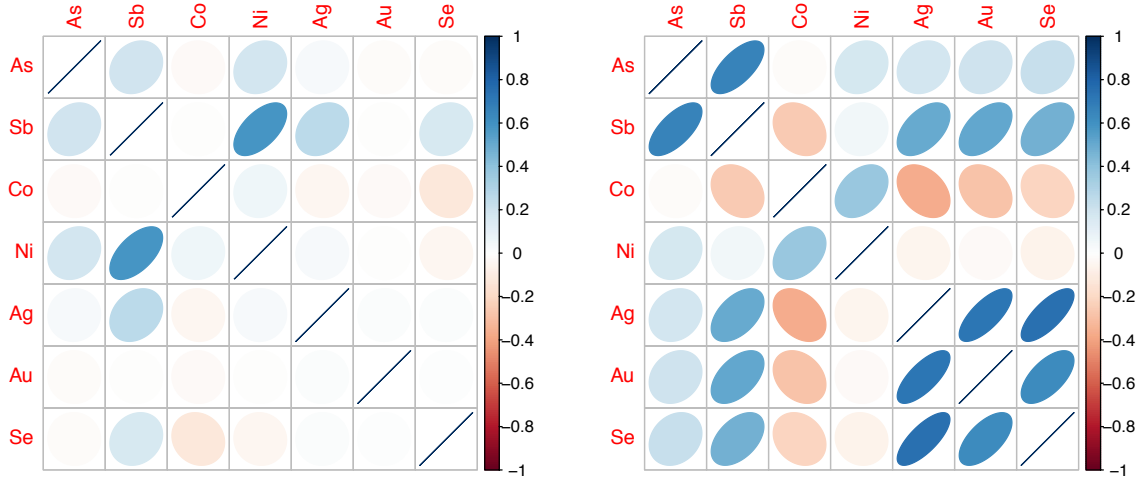

FIG. 2. A) Correlation of original trace element values. Note the colour gradient on the side marking blue for positive and red for negative correlations; B) Correlation of logarithms of original trace element values. Note the stronger correlation patterns brought out by logarithmic transformation of the original trace element data.

*Louvain method* is based on the maximization of modularity $Q$, which measures the quality a certain partitioning of a network and is defined as:

$$Q = \frac{1}{2m} \sum_{ij} \left[ A_{ij} - \frac{k_i k_j}{2m} \right] \delta(c_i, c_j)$$

where $A_{ij}$ is the weight of the link, $k_i$ and $k_j$ are weighted degrees (also known as strengths – the sums of the weights of all the links coming from that node) of the nodes $i$ and $j$, $m$ is the half sum of all the weights in the network, and $\delta(c_i, c_j)$ is delta function, which will be 1 if the nodes $i$ and $j$ belong to the same cluster $c_i$ ($c_j$). Modularity $Q$ can result in values between -1 and 1, and the larger the value, the better the partitioning of the network. This is because more links exist between the nodes of the same cluster in contrast to the links between the nodes of different clusters. Louvain algorithm includes an additional benefit in that it maximizes the partitioning of the entire network (ie level 1) but also produces alternative partitioning (level 2, level 3 etc), where modularity reaches a local maximum.

The Louvain method yielded results on two levels for our Artefacts Network: the one with 6 clusters that corresponded to the global maximum of the mentioned algorithm, and another with 10 clusters, which represented a refined picture of the initial partitioning with 6 clusters. After opting for the level with 10 (chemical) clusters in order to acquire a better resolution, we tested the significance of our results with network randomization (bootstrapping). For this, we used the resulting partitioning of our network and then randomised it, keeping only the important properties (for Artefacts Network, the weight of links). This process was repeated 1000 times, and it yielded the distribution of 1000 modularity values in our randomised network, which we then compared with the modularity value of the Artefacts Network (0.3088). The mean of the distribution of modularities of the randomized networks is 0.1012. The latter has the standard deviation of 0.0008, making the value of the original network 280 standard deviations larger then the mean of the randomized networks values (see Fig. 3a). This corresponds to the $p$ value of <0.001.

Finally, although there are other methods that can be used for determining the number of chemical clusters, we developed this one for two main reasons: a) it offers a clear criterion for obtaining the number of modules by maximising the value of modularity (unlike, for example, hierarchical clustering) and b) it gives us an option to test the significance of the obtained clustering structure with bootstrapping, by using comparison between the value of modularity and the value of randomized networks. This is the first time this method has been developed for application on archaeological data and although it differs from the average-link clustering employed by Pernicka et al [19], we observed a general consistency of our results with the outcomes of their clustering analyses (Supplementary Information, Figures S2 and S3).

*Sites Network – community structure analyses of archaeological sites.* Our second and final step (Sites Network) has archaeological sites as nodes, where links amongst them stand for pairs of artefacts from these sites that share the same chemical cluster. This relationship was established under the assumption that two artefacts belonging to the same chemical cluster could have ended up from the places of exploitation or production in two different sites through either direct or indirect contact (i.e. various types of intermediaries); we encompass both options under the term 'supply network'. The link between the archaeological sites in out Sites Network practically works in the following way: artefact A and artefact B from two different sites (nodes) belong to (chemical) cluster 1, and therefore these two sites have a link. If these two sites contain more artefacts from the same cluster, the weight of the link is larger. For example: if site $i$ contains artefacts from clusters [0,1,1,1,1,2,2,2,3] and site $j$ has artefact from clusters [0,1,1,2,2,8,9], then the weight of the link is 5 (one for each artefact of the common type). We analysed the final network with Louvain algorithm and acquired only one level with three distinctive community structures (Modules 0, 1 and 2). For the Sites Network randomisation procedure, we cut each link and randomly reconnected it to a different node while saving only the information of the degree of each node for this type of network; this procedure was repeated 1000 times. We took into consideration, for instance, that the link with weight 5 is actually 5 links. The modularity of the original network (Sites Network) is 0.276; it is also 57 standard deviations larger from the mean of the modularities of the randomized network (0.078 ± 0.004), hence confirming significance of our final network (Fig. 3b). Geographical coordinates of archaeological sites/nodes (Table S1) were used solely for illustrative purposes in this paper. The Sites Network is the final outcome of our network design, the only one whose modularity we discuss below, and which we refer to in our results and discussion.

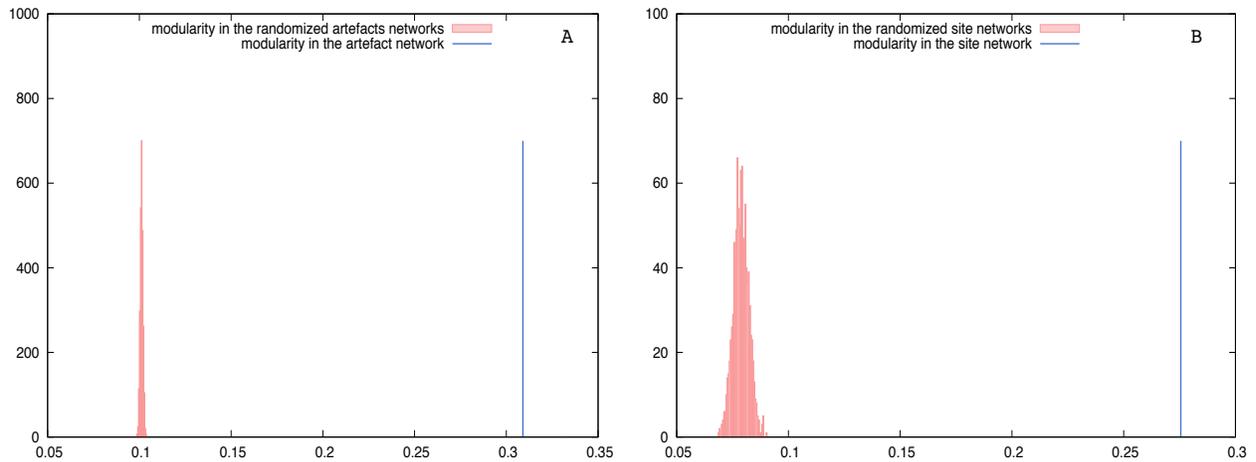

FIG. 3. A) Testing the significance of the Artefacts Network. The blue line represents the original modularity of the Artefacts Network and the red lines the distribution of modularities of randomized networks; B) Testing the significance of the Sites Network. Modularity of original network (blue line) and the modularities obtained through the randomized networks (red lines).

**The importance of archaeological sites (nodes).** We tested the importance of our nodes with three different node centrality measurements: degree centrality (based on number of links each node includes), PageRank [47] and betweenness centrality [48]. All three yielded meaningful results for determining the importance of the specific archaeological sites. The degree centrality of the node (in this case weighted degree or strength) tells us with how many other sites the observed site had some kind of communication. The PageRank takes into account how important the observed sites are. However, given that our network is not directed, these two properties appear significantly correlated (see Figure S5), and hence both presented similar results for our study. On the other hand, the betweenness centrality is defined as a number of shortest paths that go through an observed node. In order to calculate it, we defined the weights as $1/w$ or $1/w^2$, where $w$ is the weight in the original network; this procedure ensured that if there were more connections between the sites, it was easier to travel between them. Once we compared the betweenness centrality and the PageRank we observed that, barring the large difference for nodes of smaller PageRank values, the more important nodes were still more important by both measures (see Figure S6). Also, the betweenness centrality measure is not very robust and by removing only one artefact from the original input, the values change substantially, although again the more important nodes still come out the same. To conclude, using any of the importance measure yielded very similar results, which is why we give all three in Table S2. For the purpose of illustration in our maps (size of the nodes) we opted for PageRank.

The modularity analysis produced three modules with high statistical confidence, while centrality measures indicated the importance of sites in our network, reflected in the size of nodes in Figures 4-7. These three modules are interpreted as representatives of one or many copper supply networks, where the strength of links between nodes defines their membership to a module in a particular period (which may vary, see Table S2, Figure S4). This does not exclude the fact that nodes from different modules are also interconnected, rather that these links are peripheral and less strong than the strength of the links within the origin module. A more detailed account of both the nature of data and the procedure we developed is placed in the Supplementary Information, which represents an extended version of the Methods section.

3. Results and Discussion

**Modularity.** *Module 0* is spatially constrained to eastern Serbia and western Bulgaria (Figures 4 and 5a), and includes 50.5% of nodes in the total network. This is the only module that chronologically covers all seven periods, although to a variable extent. The most densely interconnected nodes occur between c. 4100 and 3700 BC, reflecting 60% of the total Module 0 assemblage, or 33 out of 55 copper-based artefacts (Figures 5a, 6e, Table S2). The most striking feature of the Module 0 structure is that the 4100-3700 BC period is predominantly represented with the Jászladány type axe-adzes, a hallmark of the Bodrogkeresztúr culture in eastern Serbia and western Bulgaria (see Fig. 7e). Almost two thirds of these axes feature in this module, most of which are stray individual finds; this in turn might explain the equal weight of the edges among the nodes observed in this module (Table S2). The pervasiveness of chemical cluster 2 indicates that copper supply network of Module 0 was likely organised around a single deposit, with occasional presence of another two clusters (no. 4 and 8) (Fig. 5d).

*Module 1* is the smallest one in the group of three, comprising 11.8% nodes that are largely aligned along the main riverine routes and their hinterlands in the northern Balkans (Fig. 5b, 6b, Table S1). It covers the c. 5500 – 4450 BC and c. 4100 – 3700 BC periods and a wide variety of artefacts, from copper minerals and slags to copper metal artefacts. It exclusively contains all metal production evidence in the entire assemblage (within c. 5000 – 4450 BC). The supremacy of cluster 2, followed by the minor presence of clusters 0, 4, and 9 (Fig. 5e) resembles largely the Module 0 supply network (Fig. 5d). More than 70% of artefacts in this module are ascribed to Periods 3 and 4 (c. 5000 - 4450 BC) (Figures 5e, 6b, 6c, Tables S1 and S2), and the spatiotemporal pattern represents a proxy for the distribution of metal producing societies traditionally identified as the Vinča culture (see Fig. 7a-c).

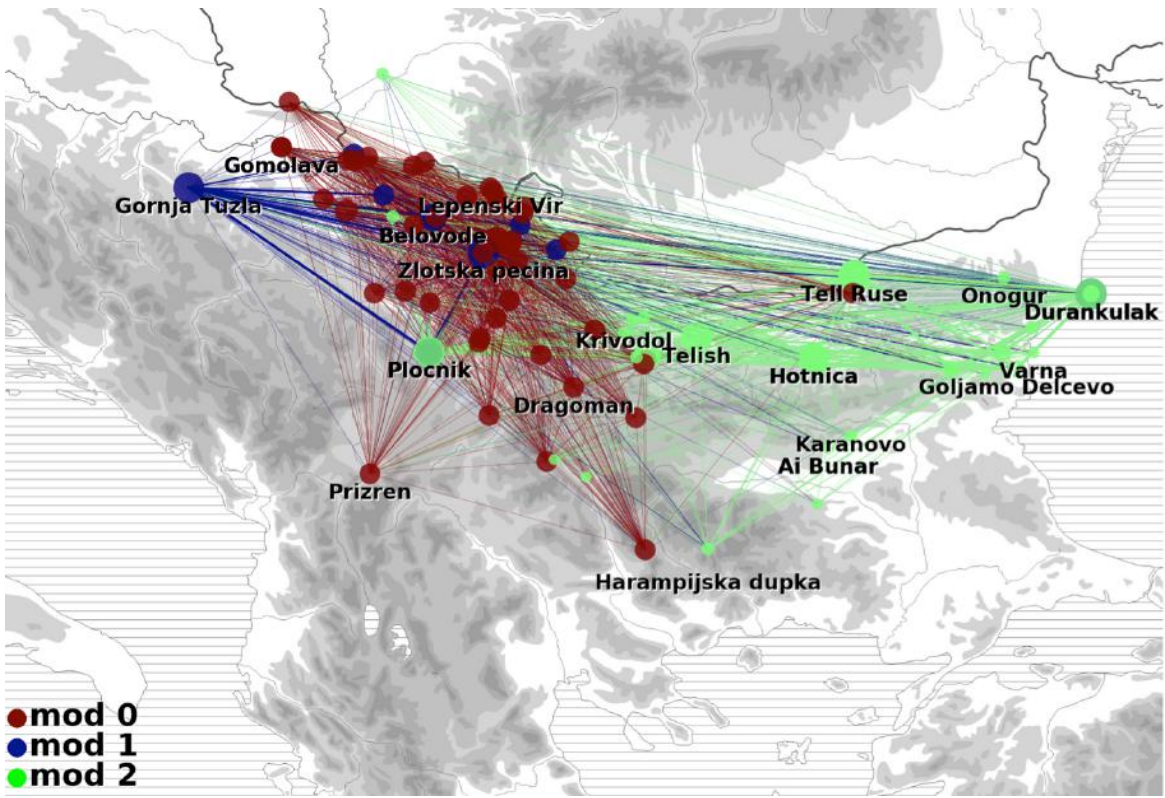

FIG. 4. Modularity analysis reveals three densely connected communities that produced and exchanged copper in the Balkans between c. 6200 BC and c. 3200 BC. Four nodes (Pločnik, Durankulak, Goljamo Delcevo, Zlotska pecina) appear in more than one module in different periods (Pločnik in Modules 0, 1 and 2, Durankulak in Modules 1 and 2, Goljamo Delcevo and Zlotska pecina in Modules 0 and 1, see Fig. S4), hence different indications of color symbol/links in this figure. The node size depends on the Page Rank value (Table S2).

*Module 2* displays the strongest spatial presence in eastern Balkans (Bulgaria, and only 4 nodes in Serbia), with 37.6% of nodes in the total network. It includes the time frame between c. 5000 BC and c. 3200 BC, with Periods 4 and 5 (c. 4600 – 4100 BC, Figures 6c, 6d) solely representing 85% of all artefacts in this module, which translates into 223 copper-based artefacts (Table S1). A great variety of copper metal artefacts dominate this module, barring a few metal casting prills. The diversified supply networks are reflected in the presence of all chemical clusters, indicating an extensive exploitation of various copper sources (Fig. 5f), in contrast to rather monopolised supply routes in Modules 0 and 1 (Figures 5d, 5e). The strongest connection between the format of this network and the established concepts of archaeological cultures is seen in the overlap of the former with the distribution of the Kodžadermen-Gumelniţa-Karanovo VI complex and the Varna culture in north-central and east Bulgaria respectively (see Figures 7c, 7d).

**Archaeological and spatiotemporal significance**. The three community structures discussed above exhibit high correlation with the known spatial and chronological dynamics of various cultural phenomena – archaeological cultures in the Balkans between the 7$^{th}$ and the 4$^{th}$ millennium BC (Fig. 7). The earliest known copper-based artefacts are included in the Module 0 assemblage, identified as copper minerals from the Early Neolithic Starčevo culture horizons at Lepenski Vir, Vlasac and Kolubara-Jaričište, dated from c. 6200 to c. 5500 BC (Figures 6a, 7a). These fall within the same module as copper minerals and beads from the early Vinča culture occupation at the sites of Pločnik (Period 2, 5500 – 5000 BC, Fig. 7a), but also Gomolava and Medvednjak (Period 3, 5000-4600 BC, Fig. 7b). Thus, communities from these sites were members of the same copper supply network throughout Periods 1-3, a notion also supported via the exclusive presence of cluster 2 at the time (Fig. 5d). Lead isotope analyses of copper minerals from these sites, which were not taken into account for our networks analyses, offer further support for the link between these artefacts by confirming their origin from the eastern Serbian sources, like Majdanpek [22, 39] (see also Fig. S2).

The presence of one node from the Vinča culture in Module 1 in Period 2 (5500-5000 BC, Belovode) might be taken as an exception to the rule, or explained as the emergence of a parallel copper supply network that develops fully in the following, Period 3 (5000-4600 BC), and that included only Vinča culture sites with the confirmed evidence for metal production (Figures 6b, 7b), as for instance Vinča, Selevac and Pločnik [21, 49, 50]. To this group we also add the site of Durankulak, located on the western Black Sea coast, with which early copper mineral and metal exchange ties with the Vinča culture metal-producing sites were known from earlier archaeological studies [19, 51]. Noteworthy is the change of modules for the site of Pločnik (from Module 0 to Module 1) in Period 3, which happens in parallel with the introduction of metalmaking activities in this prehistoric village [50].

Module 2 communities start appearing from Period 3 (5000-4600 BC) as well; two nodes, Slatino and Marica (Figures 6b, 7b), emerge in the hinterlands of the copper-rich deposits in central Bulgaria, and one of the two prominent ancient mines in the Balkans, Ai Bunar. This particularly prolific copper mine in the past is the probable source for cluster 4 artefacts (see Fig. S3), which dominates Module 2 throughout all periods (Fig. 5f).

Period 4 (4600-4450 BC) is archaeologically known as the time of the collapse of the Vinča culture (mostly northern sites), spread of the Krivodol-Salcuţa-Bubanj Hum (KSBh) I cultural complex, and the rise of two distinctive cultural phenomena in Bulgaria: Kodžadermen-Gumelniţa-Karanovo (KGK) VI and Varna cultures [52]. This situation is well exemplified in the fragmentation of copper supply networks in Figures 6c and 7c, which appear severed due to the mentioned developments.

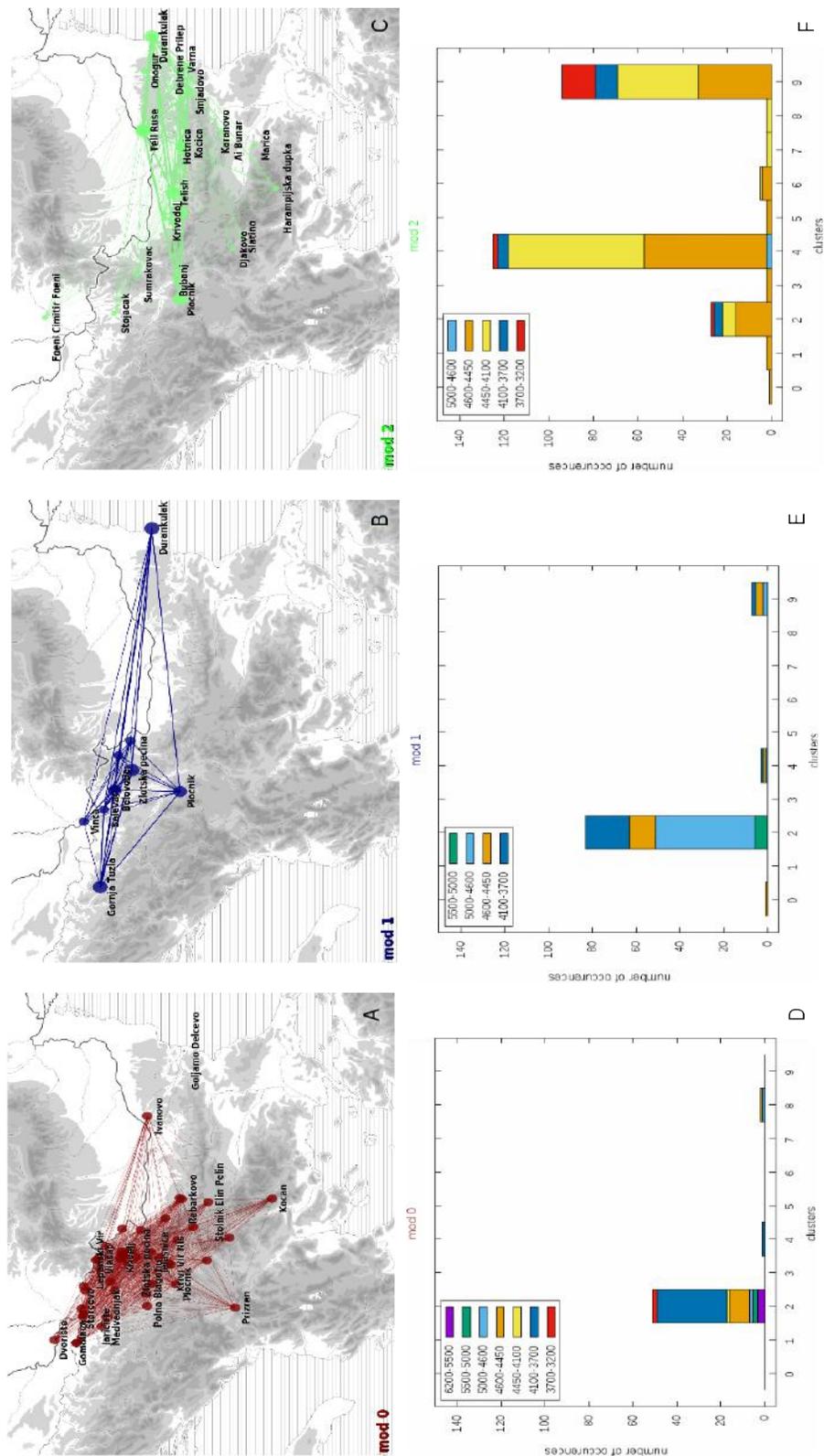

FIG. 5. Three individually presented modularity structures of copper producing and exchanging communities in the Balkans (c. 6200 – c. 3200 BC) paired with diagrams illustrating the frequency of ten different chemical clusters within each of these modules, throughout different periods. (A, D) Module 0 is represented with 50.5 % of nodes in the total network and three chemical clusters only, of which No. 2 is predominant and covers c. 6200 – 3200 BC. (B, E) Module 1 is represented with 11.8 % of all nodes and four chemical clusters. Within the chronological span of c. 5500 – 4450 BC and c. 4100 – 3700 BC, chemical cluster No. 2 is the most dominant, while clusters 0, 4 and 9 have a minor presence. (C, F) Module 2 includes 37.6% of nodes in the total network and includes all ten chemical clusters (0-9). Chronologically it covers the period between c. 5000 and c. 3200 BC, with two divisions (c. 4600 – 4450 BC and c. 4450 – 4100 BC) representing together 85% of all artefacts in this module.

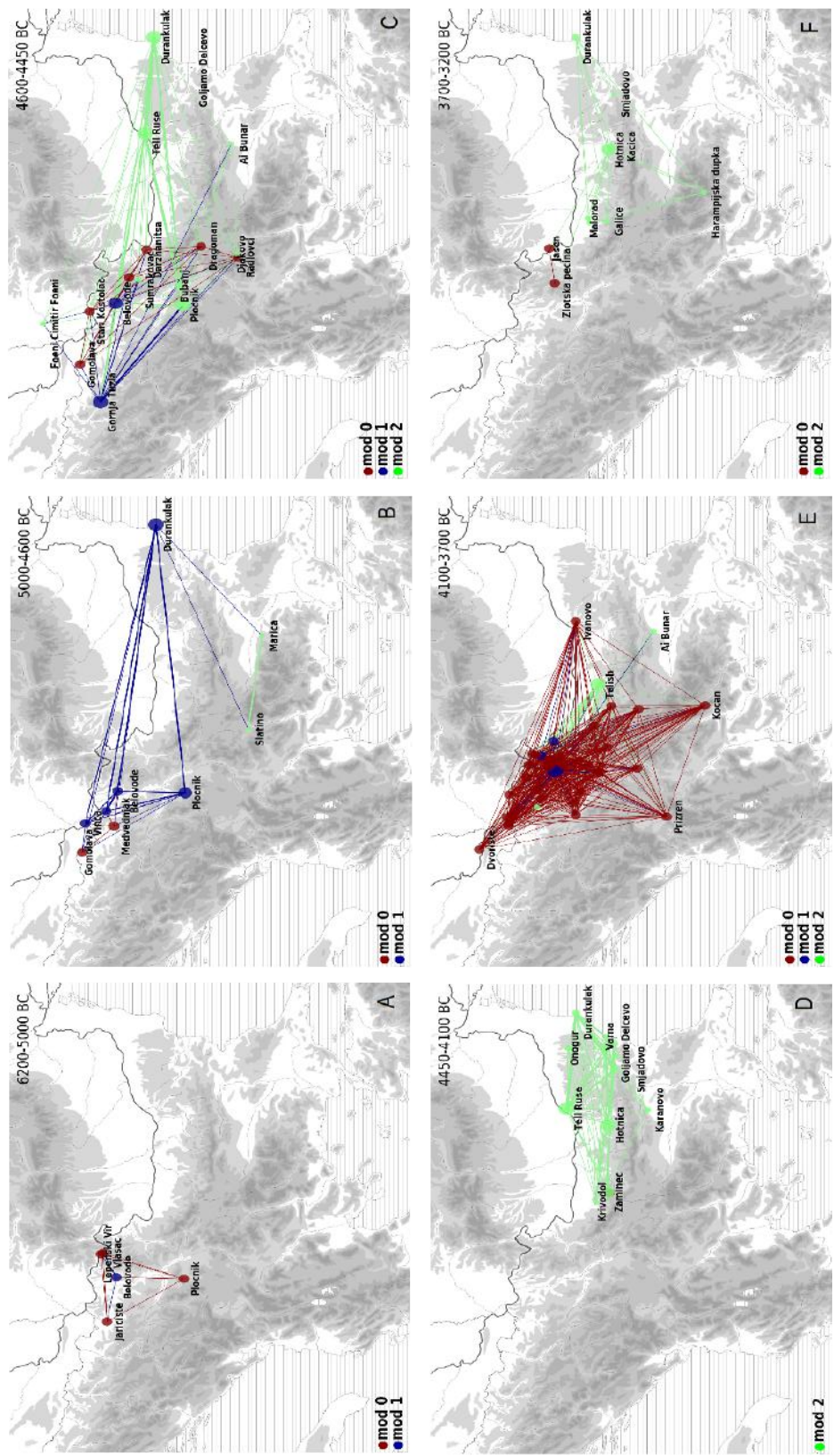

FIG. 6. Three community structures presented throughout c. 6200 – 3200 BC in the Balkans. (A) Period 6200-5000 BC (merged from 6200-5500 and 5500-5000 for presentation purposes) illustrates the early core of supply networks for copper mineral-only artefacts (Module 0) and the emergence of Module 1; (B) Period 5000-4600 BC is dominated by the supply networks of Module 1 (proxy for Vinča culture, see Fig. 7) and followed by two more nodes from Module 0 and an appearance of new copper supply network (Module 2); (C) Period 4600-4450 BC is dominated by the developing Module 2, which emerged in parallel with the slow disappearing supply regional networks of Modules 0 and 1; (D) Period 4450-4100 BC demonstrates the supremacy of Module 2 in the east Balkans (proxy for Kodžadermen-Gumelniţa-Karanovo VI cultural complex, Fig. 7); (E) Period 4100-3700 BC shows the rise of supply networks of Module 0 in central Balkans (eastern Serbia, proxy for Bodrogkeresztúr culture, Fig. 7) following the collapse of the eastern Balkan networked systems by 4100 BC; (F) Period 3700-3200 BC presents a picture of nodes scattered in eastern Balkans (Module 2) and a small cluster of Module 0 in eastern Serbia, altogether reflecting the incoherent set of available data (see Supplementary Information).

The most notable example is again the communities at the site of Pločnik, which after the collapse of the northern Vinča culture settlement, established ties with Module 2 communities, largely represented with the KGK VI complex (Fig. 7d, Table S1). These ties have already been confirmed with the lead isotope analyses that suggested Bulgarian deposits as likely sources for some copper metal implements discovered in Pločnik in its latest phases of occupation [18].

Another similar example is the site of Durankulak, which changed supply network and, similarly to Pločnik, transferred into another module (from Module 1 to Module 2, Figures 6b, 6c, 7b, 7c). Interestingly, the sites of Pločnik and Durankulak emerge as the most important in the PageRank analysis, along with Tell Ruse in the same period (4600-4450 BC, Table S2). This practically means that these nodes/sites contained most artefacts that shared chemical cluster with a number of others, all of which seem to appear only in Module 2 (Table S2). This is also the time when 8 out of 10 chemical clusters occur in the supply network of Module 2, probably implying the intensification of multiple sources exploitation (Fig. 5f), a trend archaeologically recognised earlier as the Balkan 'metal boom' [53, 54].

Module 0 sites in Period 4 (4600-4450 BC) are largely representative of the spread of KSBh I cultural complex (Fig. 7c), although there are exceptions to the rule. Period 5 (4450-4100 BC) is dominated by Module 2 supply network solely, which carries high spatial and temporal resemblance with the KGK VI cultural complex, followed by developed phases of the Varna and KSBh I cultures (Figures 6d, 7d, Table S2). These archaeological phenomena are, in terms of (copper-based) metallurgical practice and knowledge almost indistinguishable. The extensive lead isotope studies of copper metal artefacts from this period [19] suggested common use of several copper deposits across modern day Bulgaria, with distinctions only noticeable in the style of implement-making, like axes of a particular type (see Table S1). Such an 'open market' approach to acquisition of copper ores or copper metal ingots is well illustrated in Fig. 5f, which shows the presence of 6 out of 10 chemical clusters in this period. The prevalence of clusters 4 and 9 indicates clearly different extent of dependence on these potential sources (or cluster of adjacent sources with a similar chemical signature) in comparison with Modules 0 and 1, which are dominated by cluster 2 (Figures 5d, 5e).

The following period (4100-3700 BC) reveals a shift in copper supply networks from Bulgaria towards east and central Serbia (Figures 6e, 7e). This shift in copper supply networks is consistent with the known collapse of the KGK VI complex in Bulgaria and the emergence of the Bodrogkeresztúr phenomenon, confirmed as dependant on copper sources in eastern Serbia, like Majdanpek [19]. The striking dominance of cluster 2 confirms this earlier finding further, while the use of cluster 4 in this period may be matched with the remnants of the Module 2 network, presented with the sites belonging to the latest phase of KSBh IV and the rise of important settlements like Telish (Fig. 7e, Table S2). The Module 0 nodes in this period stand for finds from three hoards, carrying mixed typological characteristics of the Bodrogkeresztúr and KSBh IV phenomena (Table S1).

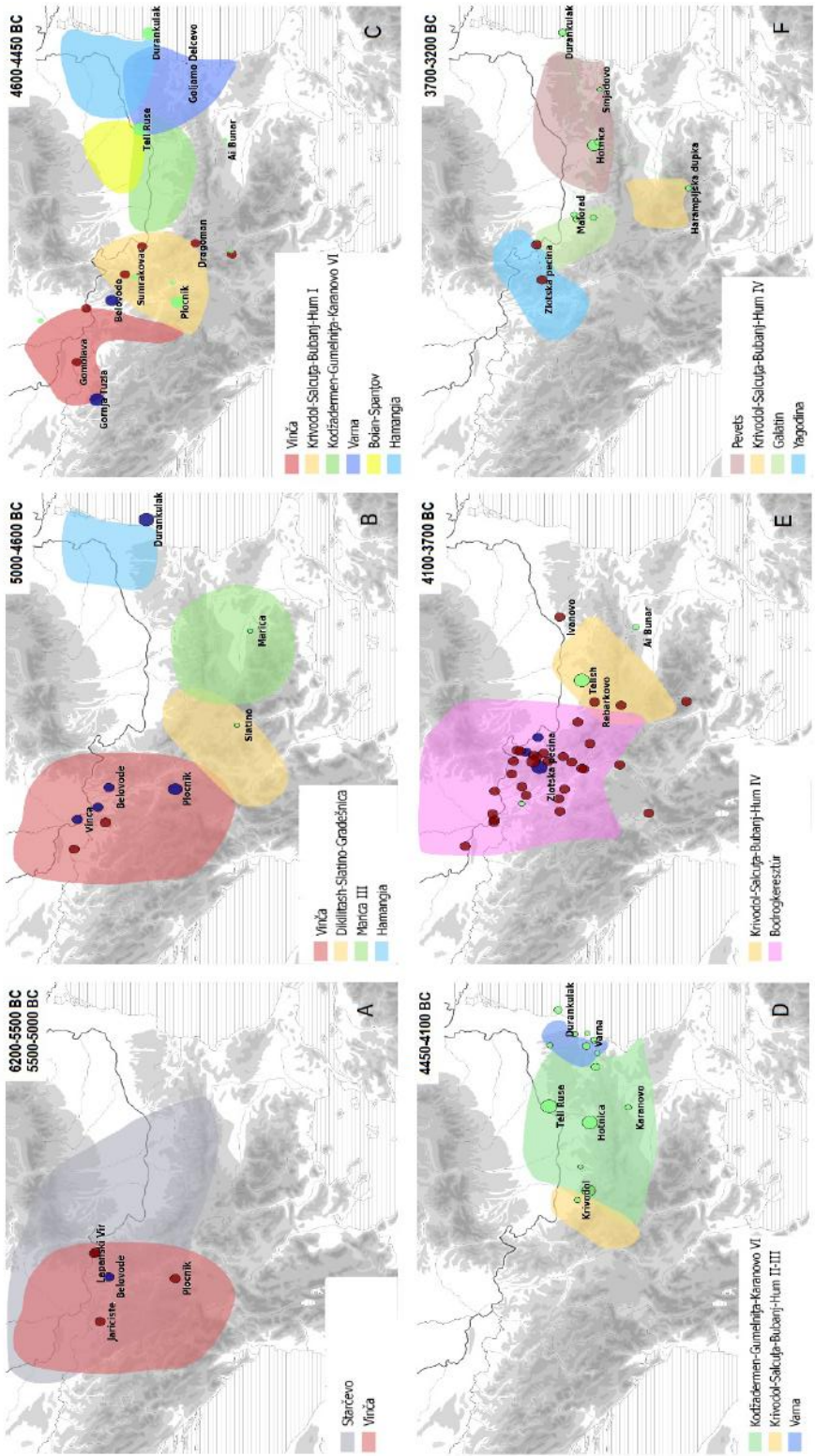

FIG. 7. Distribution of archaeological cultures / copper-using societies in the Balkans between c. 6200 and c. 3200 BC, with the most relevant sites. Note colour-coding and size of nodes consistent with the module colour (Module 0 – red, Module 1 – blue, Module 2 – green) and PageRank respectively.

The last observed period (3700-3200 BC) shows the disintegration of copper supply networks (Figures 6f, 7f); this is presented with individual or stray finds of typologically distinctive artefacts for the Proto Bronze Age, mixed with a few artefacts from the previous period [52]. The said situation is archaeologically reflected in the documented discontinuation of the 5$^{th}$ millennium BC metalmaking techniques in the Balkans, which were unique across this space, particularly in the finishing techniques for making massive copper implements, a hallmark of the mentioned millennium [39, 55, 56].

## 4. Conclusion

Here we capture properties of highly interconnected systems – community structures made of supply networks that reflect organisation of copper industry and effectively social and economic ties in the Balkans between c. 6200 and c. 3200 BC. The intensity of algorithmically calculated social interaction reveals three main groups of communities that are spatiotemporally and statistically significant (Figures 4 -7). We notice selective formation of networks ties amongst sites' populations in relation to specific regional copper sources (e.g. eastern Serbia, central Bulgaria) or communication routes (e.g. lower Danube), as well as their association with either seemingly monopolised (dominance of cluster 2, Figures 5d, 5e) or 'open-market' (all clusters included, Fig. 5f) organisation of copper supply networks across the observed periods. Noteworthy is that clusters 2 and 4 are overall coherent with the chemical signatures of the ancient mines of Majdanpek and Ai Bunar respectively (Figures S2, S3 and discussion in SI), a result consistent with previous research on metal provenancing in the Balkans [18, 19]. Importantly, the observed spatial and chronological divide of clusters 2 and 4 artefacts (Figures 5 and 6) confirms the archaeological significance of our results. It also further indicates the tendency of communities identified as archaeological cultures to maintain their own regional network of copper exploitation, production, exchange and consumption. In this light, metal recycling practices are plausible, although they would likely occur within community structures defined by our modularity research (see Supplementary Information).

The complex wiring topologies of three modules identify the evolving and collapsing past social structures, which are quantified for the first time independently from cultural, chronological and geographical attributes. These networked structures also carry strong resemblance with at least three dominant economic and social cores of copper industries in the Balkans across c. 3000 years, traditionally defined as Vinča, KGK VI & Varna, and Bodrogkeresztúr cultures through Modules 1, 2 and 0 in respective chronological order (Figures 5-7). Besides suggestive spatiotemporal patterning, such resemblance shows that algorithmically calculated community structures represent currently the most precise mathematical model for identifying archaeological phenomena, or what appears to show here as the selective formation of past social ties across the observed period. The dynamics of copper exploitation, production and consumption practices in this case reflect closely the dynamics of recorded social interaction within the observed time and region. Although we are not suggesting that metallurgy-related practices were the sole factor in defining these interactions, such as collapses or rises of cultural complexes, our research indicates that they must have been sufficiently powerful to play a major role in their shaping.

The results of our analysis indicate that: a) the obtained community structures exhibit patterns of cooperation and complex human interaction that derive from shared economic interests, cultural values [57], beliefs and probably common communication system; b) these patterns were obtained independently of inputs like cultural, chronological or geographical attribution of data and c) the consistency of gained structures with the distribution of archaeological cultures illustrates that foundations of our method are reliable. Its resolution and verified probability demonstrate capacity for revealing a significantly nuanced picture of past social interaction, in contrast to the traditional methods. We therefore suggest that community structure property of networked systems represents a compelling tool for an independent evaluation of the archaeological record.

The obtained high-resolution picture of dynamics of networked systems in the past provides a model that stands in stark contrast to the static nature of determining patterns of social interaction based on accumulation of seemingly similar material traits in the archaeological record. A good example for the high-resolution result is Pločnik, a site that changed modules throughout three distinctive periods (Figures 6a, 6b, 6c). Looking closely at the nature of these changes, it can be seen that they were triggered at the time of the known developments: emergence of metallurgical activities (c. 5000 BC, Module 0 to 1) and collapse of the majority of networked structure occupied by communities labelled as the Vinča culture (c. 4600 BC, Module 1 to 2). Our method can, therefore, reveal connections that traditional systematics was not able to and provide insight into the dynamic nature of (re)forging relations in the altering climate of social and economic evolution of past societies. Equally so, complex networks analyses allow us to observe the spread of technological innovations, like metallurgy, through the lens of cooperative networks and processes involved in their formation and maintenance. It would be relevant for this study to compare our results with any demographic changes at the time. Although a few innovative studies have addressed this matter earlier [58, 59], along with the exemplary success related to networks analyses in the pre-Hispanic US Southwest [28], there have not yet been attempts to integrate metallurgical and population data for the $7^{th}$ - $4^{th}$ millennium BC Balkans.

We select two key arguments to highlight the novelty of our model for studying community structures in the human past record. One is that it is based on a variable independent of any archaeological and spatiotemporal information, yet provides archaeologically and spatiotemporally significant results. The second is that a study of community structure property of networked systems in the past produces coherent models of human interaction and cooperation that can be evaluated independently of established archaeological systematics. Despite the imperfect social signal extracted from the archaeological record, our method provides important new insights into the evolution of the world's earliest copper supply network and establishes a widely applicable model for exploring technological, economic and social phenomena in human past, anywhere.

## 5. Acknowledgements

We thank the many individuals and institutions that contributed data and access to materials, including constructive comments during the study development. We are particularly grateful to M. Charlton, B. Roberts, E. Crema, J. Pendić, A. Clarke and S. Shennan for their kind help and suggestions. We are indebted to Prof. M. Milinkovic (University of Belgrade) for permission to use the base map of the Balkans. J.G. and M.R. acknowledge the financial support from McDonald Institute for Archaeological Research, University of Cambridge, UK's Arts and Humanities Research Council (project AH/J001406/1) and FWO Research Foundation - Flanders.

59. PORČIĆ, M., BLAGOJEVIĆ, T. & STEFANOVIĆ, S. (2016) Demography of the Early Neolithic Population in Central Balkans: Population Dynamics Reconstruction Using Summed Radiocarbon Probability Distributions. *PLoS ONE,* **11,** e0160832.

# Supplementary Information

## *Community structure of copper supply networks in the prehistoric Balkans: An independent evaluation of the archaeological record from the 7th to the 4th millennium BC*

M. Radivojević[*]

[1]McDonald Institute for Archaeological Research, University of Cambridge, Downing Street, CB2 3ER Cambridge, UK
[*]Corresponding author: mr664@cam.ac.uk

J. Grujić[*]
[2]Vrije University Brussels, Artificial Intelligence Lab, Pleinlaan 2, Building G, 10th floor & MLG, Département d'Informatique, Université libre de Bruxelles, 1050 Brussels, Belgium
[*]Corresponding author: jgrujic@vub.ac.be

**Data.** For networks analyses we used published and unpublished sets of compositional data for 410 copper-based objects under consideration, spanning c. 6200 to c. 3200 BC. These were assembled from publications of Pernicka et al. [1, 2], Radivojević et al. [3], Radivojević [4] and UK's AHRC-funded "Rise of Metallurgy in Eurasia" project (hosted by the UCL Institute of Archaeology, no. AH/J001406/1)[5]. All data originate from the analytical set up of a single laboratory, Centre for Archaeometry in Mannheim, Germany, led by Professor Ernst Pernicka (DOI: 10.17863/CAM.9599). The data from 410 objects are presented with a unique laboratory number (given by Centre for Archaeometry) and include the following types of materials (Table S1): copper mineral (30 in total), mineral ornament (17 in total), production evidence (smelting/casting, 22 in total), metal ornament (99 in total), and metal implement (242 in total). The metal implement type some instances contained information on the type of axe, included in a separate column (no data labelled as: *unk*). Besides a unique geographical location (given as latitude and longitude in degrees), sites are ascribed the following regional codes: SRB (Serbia), W (West Bulgaria), THR (Thrace), RHD (Rhodope), NC (North-central Bulgaria), NE (North-east Bulgaria) and BSC (Black Sea Coast). The dataset is available under: https://doi.org/10.17863/CAM.9599.

Chronological and cultural attribution of studied materials was ascribed based on available relative and absolute dating in the area under consideration [1-3, 6-26]. Seven cultural periods were designated for this study: Early/Middle Neolithic (Period 1, 6200-5500 BC), Late Neolithic (Period 2, 5500-5000 BC), Early Chalcolithic (Period 3, 5000-4600 BC), Middle Chalcolithic (Period 4, 4600-4450 BC), Late Chalcolithic (Period 5, 4450-4100 BC), Final Chalcolithic (Period 6, 4100-3700 BC) and Proto Bronze Age (Period 7, 3700-3200 BC). We would like to emphasise that there is not a general consensus on the relative vs. absolute chronology of existing cultural phenomena observed here amongst (Balkan) archaeologists; thus, the entries on chronology in Tables S1 and S3 should be taken as tentative interpretations based on the latest chronological update in the field.

*Period 1. Early/Middle Neolithic (EN, 6200-5500 BC).* This period is represented only with the finds belonging to the Neolithic occupation of sites in eastern and western Serbia (proto/Starčevo culture) (see Figure 7a, Table S3). The use of malachite at the time of the introduction of agriculture or domestication is not uncommon, and similar examples have been documented in the Near East.

However, the minerals listed here are unique since their function remained unknown, although the provenance analyses indicate their origin from local, eastern Serbian sources [18].

| Period | C14 dates | Vojvodina | Central Balkans | West Bulgaria | South Bulgaria | Muntenia | North-east Bulgaria | Black Sea Coast (west) |
|---|---|---|---|---|---|---|---|---|
| Proto Bronze Age | 3200 | Boleráz | Cernavoda III | **Galatin** | **Yagodina** | Cernavoda III | **Usatovo** | Cernavoda I |
| Final Chalcolithic | 3700 | **Bodrogkeresztúr** | **Salcuţa IV** **KSBh** | | | Cernavoda I | **Cernavoda I/Pevets** | |
| Late Chalcolithic | 4100 | **Tiszapolgár / KSBh** | | **Krivodol-Salcuţa-Bubanj hum (KSBh)** | **Karanovo VI** | **Kodžadermen-Gumelniţa-Karanovo VI** **Boian-Spanţov** | **Kodžadermen-Gumelniţa-Karanovo VI** | **Varna III** **Varna I** **Varna II** **Hamangia IV** |
| Middle Chalcolithic | 4450 | **Vinča D** | | | Marica IV | | | |
| Early Chalcolithic | 4600 | **Vinča D** **Vinča C** | | **Gradešnica** **Dikilitash-Slatino** | **Marica III-Karanovo V** | Boian-Vidra | Poljanica | **Sava / Hamangia III** |
| Late Neolithic | 5000 | **Vinča B** **Vinča A** | | Kurilo/Akropotamos Topolnica | Karanovo IV Karanovo III | Boian III | Hotnica | **Hamangia II** **Hamangia I** |
| Early Neolithic | 5500 6200 | **Starčevo** **Lepenski Vir III** | | | | | | |

**Table S3. Relative and absolute chronology of malachite and metal-using cultures in the Balkans, 7th – 4th mill BC. Green font stands for using copper minerals (e.g. malachite beads), red for metallurgical materials (e.g. metal artefacts, slags). The shaded fields indicate the periods and regions covered in this study.**

*Period 2. Late Neolithic (LN, 5500-5000 BC).* This period is linked with the emergence of archaeological cultures that would grow into large metal producing and consuming phenomena (like Vinča in Serbia or Karanovo in Bulgaria) during the 5th millennium BC [27]. While Vinča culture occupied most of the central Balkans, the Karanovo phenomenon emerged in central Bulgaria and expanded significantly in the second half of the 5th millennium BC, including territories from the Black Sea coast to Thrace. With settlements rising on river plateaux across the region, exceptional craftsmanship in pottery and stone industry started to develop. Copper minerals and malachite beads found in settlements and cemeteries at the time became more numerous, although no thermal treatment has been recorded prior to c. 5000 BC. In eastern Serbia, the Vinča culture communities commenced mining activities in the currently earliest known copper mine, Rudna Glava [28]. Although no metal artefacts from this period (or later) were found to compositionally and isotopically match this source, there were other mines in otherwise copper-abundant region of eastern Serbia, such as Majdanpek, which were particularly prolific during the 5th millennium BC [1].

*Period 3. Early Chalcolithic (EC, 5000-4600 BC).* The start of the copper smelting activities is set around 5000 BC [3], which corresponds with the earliest known metal artefacts appearing in the Vinča culture site of Pločnik in south Serbia, followed by similar finds along the Black Sea Coast and in south Bulgaria [29]. Settlements grew in size, particularly along the lower Danube, which was probably the easiest and quickest means of transport for the emerging long-distance trade of prestigious commodities, such as spondylus, obsidian, malachite beads and metal artefacts, amongst others [30, 31]. Many of these were found in the first organised cemeteries at the time, probably designating high status of buried individuals (i.e. Durankulak). The 'metal effect' is seen in the occurrence of graphite-painted decorations on pottery at the time, possibly imitating one of the most desired materials of the 5$^{th}$ millennium BC in the Balkans. Towards the very end of this period, archaeologists have recorded the first mining activities in Bulgaria, at the site of Ai Bunar, started by the bearers of the Marica culture [32, 33] (Table S3). It grew to be the most important source of eastern Balkan region throughout the later 5$^{th}$ millennium BC.

*Period 4. Middle Chalcolithic (MC, 4600-4450 BC).* This period is difficult to separate out from what appears to be an uninterrupted evolution of metal making cultures in eastern Balkans (Bulgaria) and slow disintegration of the Vinča culture in Serbia. It is generally characterised with the rise of two large cultural complexes and one culture in northeastern Bulgaria. While northern Vinča culture sites were rapidly being abandoned and conflagrated, a few southern ones (like Pločnik) continued to live until the very end of the Vinča culture in south Serbia (c. 4450 BC). Some scholars argued that it was the late Vinča culture in this region and Gradešnica in west Bulgaria that gave impetus for the formation of the Krivodol-Salcuţa-Bubanj Hum (KSBh) I cultural complex [21]. The other large cultural complex was formed by the merging of Marica, Karanovo V and Boian Spanţov cultures in south Bulgaria and Muntenia, and is known under the name of Kodžadermen-Gumelniţa-Karanovo (KGK) VI (Table S3). Varna culture, named after the eponymous burial site with the world's earliest gold objects, occupied the (western) Black Sea coast. This is the time when large tell-sites dotted both riverbanks along the lower Danube, but also other regions in Bulgaria (like Karanovo tell-settlement, for instance). Metal production enters its peak production, where diversification in copper hammer-axe design was most likely due to communities seeking for a personal stamp in then fast-expanding metalmaking industry.

*Period 5. Late Chalcolithic (LC, 4450-4100 BC).* While in some parts of Bulgaria the transition from the previous period into this one is hardly recognisable in material culture, the Late Chalcolithic period has been supported with absolute dates. The material culture and domestic architecture are developing in this period together with the extensive burial evidence for the rising wealth of individuals, and hence a potential social stratification and emergence of an elite [34]. A rapid climate change towards the end of this period is seen as the major cause of disappearance of any record of the communities in the east and central Bulgaria. The disintegration of the communities seemingly started with the coming of the steppe population, although the complete cultural caesura must have been a combination of several factors [24]. In the west, KSBh cultural complex spreads over a vast space between Oltenia and the Aegean (Thassos). The material expression and settling habits differ from the developments in the east, with settlements mostly established at higher altitudes or caves. In Serbia, this cultural complex borders with the Tiszapolgár culture (Table S3).

*Period 6. Final Chalcolithic (FC, 4100-3700 BC).* This period was characterised by the shift in metal-making industry towards the west of the observed area. Metallurgy intensified in the KSBh IV cultural complex, potentially due to the decline of the Thracian mining centres [21]. Evolving domestic architecture, settlements established on inaccessible paths, and innovations in pottery making were all part of this new phase of the KSBh cultural complex evolution. The mining and metal production was revived in eastern Serbia, particularly with the massive production of Jászladány type hammer axes, related isotopically to the Majdanpek mine, and culturally to the Bodrogkeresztúr culture [2]. This culture emerges east of the Tisza river, with sites dotted along its lowlands and into the Serbian Banat [15]; its southern spread is a matter of content, however, the spread of the Jászladány hammer axes indicates strong social and economic ties with area south of Danube. Gold objects occur for the first time in this part of the Balkans.

*Period 7. Proto Bronze Age (PB, 3700-3200 BC).* This period saw the final disintegration of all cultural complexes formed during the 5$^{th}$ millennium BC. Small-scale settlements with rare metal artefacts are recorded throughout Bulgaria, with new metal tools, like daggers, making the appearance for the first time, presumably echoing the Eurasian Steppe influence.

Each node in our network was followed by the designated time-period in our analyses in order to clarify which occupational horizon within a site (node) yielded which type of artefacts. Barring seven exceptions (see Table S1), all sites (or nodes) were ascribed a relative cultural affiliation based on the current state of research.

**Community structure (modularity) analysis**. Our network was built in two discrete steps: 1) we grouped the data in ten distinctive chemical clusters (*Artefacts Network*); 2) placed a connector between the sites that contain pairs of artefacts from the same cluster and analysed the final network for community structures (*Sites Network*). In both steps we used the Louvain algorithm [35] to obtain community structures (modules) and bootstrapping to test the significance of gained results.

**Artefacts Network – clustering the copper objects by trace element chemistry.** For each of 410 artefacts (Table S1) we used the readings for the following seven trace elements: arsenic (As), antimony (Sb), cobalt (Co), nickel (Ni), silver (Ag), gold (Au), and selenium (Se), since they are the ones that are commonly thought to survive the hot temperature treatment from the copper ore to the copper metal in our case [36-38]. We therefore extracted only these values (presented in Table S1) and then performed the following course of actions that led to obtaining the number of chemical clusters in our dataset:

1) transforming several compositional readings in our dataset with zero (0) value into a small positive number (0.0001); this number was smaller than the detection limit of any of the analysed elements;

2) calculating logarithms of all 7 trace elements;

3) running principal component analyses of the logged values and obtaining principal component scores;

4) determining Euclidean distance between all pairs of artefacts;

5) designing the Artefacts network with artefacts as nodes and links defined as $1/d^2$ (*d* = Euclidean distance), and

6) obtaining the number of chemical clusters after conducting modularity analyses with the Louvain algorithm.

To rationalise this sequence, we will start with justifying the modularity approach to chemical clustering. Theoretically, the goal of chemical clustering is to detect groups of copper artefacts whose compositional signature (a string of 7 trace elements) is more similar within a group (or a cluster) than with compositional signature of copper artefacts - members of other groups (or clusters). In other words, the links that join copper artefacts of the same chemical cluster are based on compositional similarity, and they are comparatively stronger within a cluster of chemically similar artefacts than the links connecting these artefacts to other clusters. Since this compares closely to the definition of network modularity [39, 40], we designed the cluster analyses based on the principles of community structure research in networks. There are other methods that can be used for determining the number of chemical clustering, however, we developed this one for two main reasons:

1) it offers a clear criterion for obtaining the number of modules by maximising the value of modularity (unlike, for example, hierarchical clustering);
2) it gives us an option to test the significance of the obtained clustering structure with bootstrapping, by using comparison between the value of modularity and the value of randomized networks.

Hence, the nodes of our network for obtaining chemical clusters were artefacts, while we defined links using Euclidean distance of the vectors of transformed trace element values. Namely, calculating Euclidean distance with the original trace element values proved challenging for two reasons: a) they showed lognormal, instead of Gaussian distribution in our case (Fig. 1) and b) they were correlated to begin with (Fig. 2a). Starting with the former, the lognormal distribution of our data indicated that small values are predominant (Fig. 1), and computing distances between the original data would lead to losing information on variation in smaller values. For instance, the difference between the values of 0.001 and 0.002 would make much smaller contribution in comparison to the difference between the values of 100 and 101. Hence, in order to account for these variations on the same scale, or same relative differences, we transformed the original values into logarithms. The logarithms of original data brought out clearly the correlations between chemical elements, like Sb and As, Au, Ag and Se, or Sb with Ag/Au/Se (Fig. 2b). This took us to acknowledging a particular (mathematical) property of compositional datasets, known as the constant-sum constraint (CSC), which refers to a constant sum of 1 or 100% for all variables in a measured sample [41, 42]. It means that individual variables in the compositional data do not vary independently – i.e. if one variable decreases, the proportion of the remaining must increase. Such an induced correlation may easily hinder the true relationships among variables (in our case trace elements), which is why the next step in our data processing was to eliminate these correlations. For this, we ran principal component analysis (PCA), a statistical procedure used to reduce the dimensionality of a dataset consisting of a large number of interrelated variables, while retaining the variation present in the dataset. The output are uncorrelated variables (principal components), ordered in a way that the first few keep most of the variation present in all of the original variables [43]. The PCA is the same procedure as eigenvalues decompositions from linear algebra. The PCA removed these correlations (Fig. S1), preparing the output, now calculated as principal component scores (Table S1), for network analysis. The logarithmic transformation, PCA, and visualisation in Figures 1, 2 and S1 were all computed in R (we used the corrplot library for correlations in these figures).

The straight approach to PCA with original compositional data has already been known as fraught with difficulties for the reasons mentioned above [41, 42]. Aitchison [41] proposed a way around these constraints by arguing that the best way to compute principal components out of restricted types of data (e.g. in allometry, or compositional data) is to use logarithms of the original data. This supports the treatment of our original data, although it was also in our case evident as a necessity from lognormal distribution (Fig. 1). A disadvantage of his approach was in that it could not handle zeros (0), which in our case was about to lead to losing a small handful of objects where particular trace elements were not detected (or were below the detection limit of the analytical instrument). An alternative, however, was to replace zero values with a small positive number, which is what we did before transforming the original values into logarithms. Our small positive number was smaller than the detection limit of any of the analysed elements (0.0001), as mentioned above.

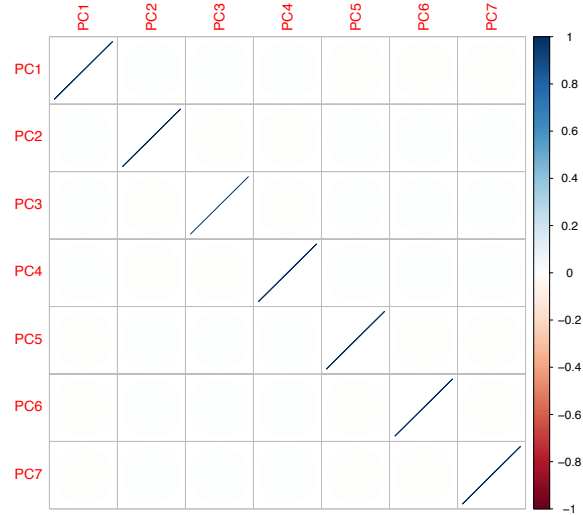

**Figure S1. The principal component analysis yielded the uncorrelated variables (compare with Figures 1 and 2).**

In the following step, the principal component scores (Table S1) were used to calculate the Euclidean distance between all pairs of artefacts. For this, we followed the rationale below: if $\vec{a}$ is a principal component vector of one artefact and $\vec{b}$ is a principal component vector of another artefact, the distance between the two artefacts will be defined as Euclidean distance between these two vectors as:

$$d(\vec{a},\vec{b}) = \sqrt{\sum_{i=1}^{n}(a_i - b_i)^2}$$

Thus, the network we formed has artefacts as nodes and links defined as $1/d^2$ ($d$ = Euclidean distance). The number of clusters was obtained with the Louvain algorithm [35]. We used the original implementation of the code written in C++ by E. Lefebvre, and later adapted by J.-L. Guillaume; it is also freely available for download on https://sites.google.com/site/findcommunities/ (the current version is maintained on  https://sourceforge.net/projects/louvain/).

*Louvain method* is based on the maximization of modularity $Q$, which measures the quality a certain partitioning of a network and is defined as:

$$Q = \frac{1}{2m}\sum_{ij}\left[A_{ij} - \frac{k_i k_j}{2m}\right]\delta(c_i, c_j)$$

where $A_{ij}$ is the weight of the link, $k_i$ and $k_j$ are weighted degrees (also known as strengths – the sums of the weights of all the links coming from that node) of the nodes $i$ and $j$, $m$ is the half sum of all the weights in the network, and $\delta(c_i,c_j)$ is delta function, which will be 1 if the nodes $i$ and $j$ belong to the same cluster $c_i$ ($c_j$). Modularity $Q$ can result in values between -1 and 1, and the larger the value, the better the partitioning of the network. This is because more links exist between the nodes of the same cluster in contrast to the links between the nodes of different clusters. Louvain algorithm includes an additional benefit in that it maximizes the partitioning of the entire network (ie level 1) but also produces alternative partitioning (level 2, level 3 etc), where modularity reaches a local maximum. Out of two levels of results, one with 6, and the other with 10 clusters, we opted for 10 and tested the significance of our results with network randomization (bootstrapping).

*Network randomization (bootstrapping)*. We performed the bootstrapping in the following way: we used the obtained partitioning of the network and then randomized it, keeping only the properties that were important (here we only preserved the weights of the links, but we shuffle the nodes they were connected to). The result was the unique partitioning of the randomized network and the corresponding modularity value. This process was repeated 1000 times, and it yielded the distribution of 1000 modularity values in our randomized network, which we then compared with the modularity value of our Artefacts Network (Fig. 3a). The calculations were run in Python using the code for the Louvain algorithm (written in C++) for obtaining the modularity values. Histograms in Fig. 3 in the main text were produced in Gnuplot.

The modularity of the Artefacts Network is 0.3088 and the mean of the distribution of modularities of the randomized networks is 0.1012. The latter has the standard deviation of 0.0008, making the value of the original network 280 standard deviations larger then the mean of the randomized networks values (see Figure 3a). This corresponds to the $p$ value of <0.001, since we randomized the network 1000 times over.

*Clustering method – consistency with the previous research.*

In the previous study on the provenance of the $5^{th}$ millennium BC Balkan copper metallurgy, Pernicka et al [1, 2] conducted average-link analyses (a type of hierarchical clustering) in order to group more than 300 copper artefacts into cohesive clusters. They initially transformed the trace element concentration of As, Sb, Ag, Co, Ni, Au, and Se into logarithms and then applied the average-link cluster analysis with Euclidean distances using the SAS (Statistical Analysis Software) program package. This program uses the cubic clustering criterion [36, 44] as the parameter for determining the optimum number of clusters, which is how Pernicka et al. [1, 2] arrived to defining nine chemical clusters in their research. They then used discriminant analysis to calculate the probability of each sample to belong to the cluster it was assigned to with the average-link procedure, and applied the 50% rule: where cases (objects) had less than 50% probability of belonging to the assigned cluster, they were re-assigned to the cluster they had the highest probability for membership.

In order to check the consistency of our clustering method (modularity) with the one described above, we tested the data from our two largest clusters, cluster 2 and cluster 4 (Table S1), against the trace element patterns of the two most prolific prehistoric copper mines in the Balkans, Majdanpek and Ai Bunar (Figures S2 and S3). Namely, Pernicka et al. [2, 117, Fig. 20] managed to identify the chemical correlation between the Majdanpek mine and their cluster 2 (58 artefacts), and the Ai Bunar mine and their cluster 3 (43 artefacts), hence providing support for the argument that these two mines/copper deposits were exploited to make the observed sets of copper artefacts from the $5^{th}$ millennium BC Balkans. We performed a similar test by plotting the trace element values of our cluster 2 artefacts (161 objects) with the trace element signature of Majdanpek (Fig. S2), and the trace element values of our cluster 4 artefacts (129 objects) with the trace element signature of Ai Bunar (Fig. S3). We chose these clusters since the former relates mostly to sites in Serbia and western Bulgaria, while the latter shows similar preferred associations with the sites in central and east Bulgaria. Also, these clusters (2 and 4) largely represented expanded versions of Pernicka et al.'s clusters 2 and 3 respectively; we were not, however, expecting the exact overlap between these given that we were working with a larger dataset than these authors.

The plot on figure S2 shows a general consistency of cluster 2 artefacts with the Majdanpek ore field (grey), with the notable exception of three samples in total (labels: MA-071499, L354, L355, see Table S1). While Ni and Ag values in the Majdanpek ore and cluster 2 artefacts appear most correlated in Fig. S2, the greatest fluctuations are noticed in the Sb, Co and Au values. The plot on figure S3 also presents a tight pattern of cluster 4 artefacts matching closely the trace element pattern of Ai Bunar ores (grey field). The trace element values in this plot are highly correlated, barring As and Sb readings.

Chemical fluctuations can be explained with several factors, both from the perspective of designated ore fields or the nature of artefacts making. Namely, when it comes to potential chemical variability in the ore fields, noteworthy is that the grey (mine) patterns in figures S2 and S3 stand for the 10$^{th}$ and 90$^{th}$ percentile of the maximum and minimum recorded trace element values for Majdanpek and Ai Bunar. Although it does not mean that the grey patterns are incorrect, there is always a possibility that the sample size representing this ore field was not representative to begin with.

Speaking of the chemical fluctuation of trace element patterns of artefacts against the original ore background, the lower readings of As and Sb in copper artefacts (in fig. S3 and partly in fig. S2) may imply the possibility of loss during metal extraction or recycling, particularly since the former has been known as volatile. The extent of volatility of As during arsenical copper recycling has been hotly debated in archaeology and archaeometallurgy, with discussions mostly concentrating on the redox conditions of the (s)melt and the compositional threshold below which As in copper becomes less volatile [45, 46]. In this light, and given that we are addressing here traces of both As and Sb (in ppm, not in percentages), we propose the recycling hypothesis only as an assumption that needs further probing. If the loss of As and Sb was indeed related to recycling in our cases, then such practice must have occurred within regionally (and potentially culturally) defined spaces. This conclusion follows neatly our modularity research, and is also addressed in the main manuscript.

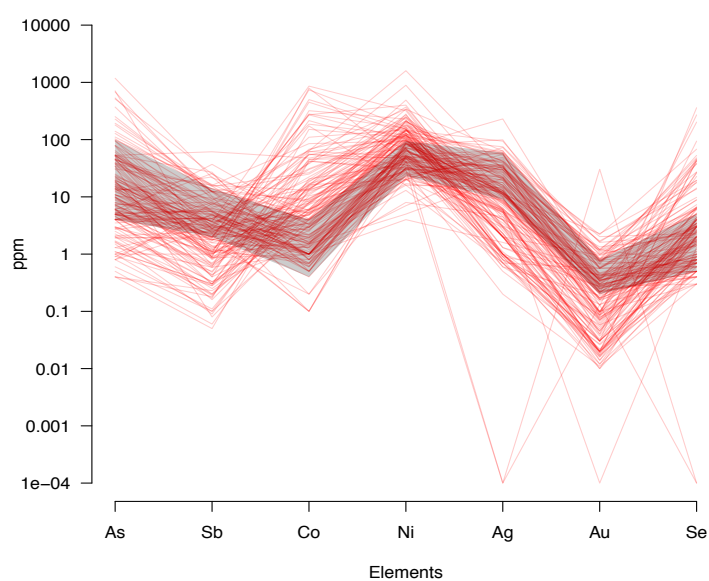

**Figure S2. Trace element signatures of 161 copper artefacts belonging to cluster 2 (lines) plotted against the trace element signature of Majdanpek (grey field), a copper mine in eastern Serbia.**

The third potential explanation for the observed chemical fluctuations is that both clusters 2 and 4 reflect chemical signatures of several deposits adjacent to Majdanpek and Ai Bunar respectively. This is not improbable given that nowadays the preserved prehistoric mining commonly represents copper deposits that survived the later exploitation (and hence destruction) as not economically feasible investments in modern terms. While Ai Bunar might represent such a case, the exploitation of Majdanpek has only been confirmed through provenance analyses thus far [1], and not through verified traces of prehistoric exploitation beyond a few chronologically indistinctive grooved hammer-stones kept in the Mining Museum in Majdanpek in Serbia. Thus, the best-case scenario for the surviving ancient mining is the poor ore quality, which may provide some grounds to presume that our two prolific mining sites in Serbia and Bulgaria may be only reflecting the less rich remnants of the actual copper mineral vein that had been mined in their vicinity.

Chronology of the plotted artefacts may also help understand the chemical fluctuations. Cluster 2 is dominated by copper artefacts from two distinctive chronological 'block periods': 5500-4450 BC and 4100-3700 BC, while cluster 4 includes mainly artefacts from 4450-4100 BC. The fluctuating pattern of cluster 2 artefacts may indicate the use of different ore sources in 5500-4450 BC and 4100-3700 BC respectively, although regionally constrained to eastern Serbia. On the other hand, the tight pattern of cluster 4 may indicate exploitation of a source in the vicinity of Ai Bunar with lower As and Sb content, or Ai Bunar itself followed by extensive recycling that took place within the constrains of the cultural / social boundaries of the KGK VI and related cultural complexes. All options presented here will be addressed in detail in future research.

Overall, figures S2 and S3 exhibit noticeable correlation of cluster 2 and cluster 4 artefacts with Majdanpek and Ai Bunar. Our clustering method shows good consistency with the cluster analyses of Pernicka et al. [1, 2], which along with bootstrapping (Fig. 3a), verifies its reliability.

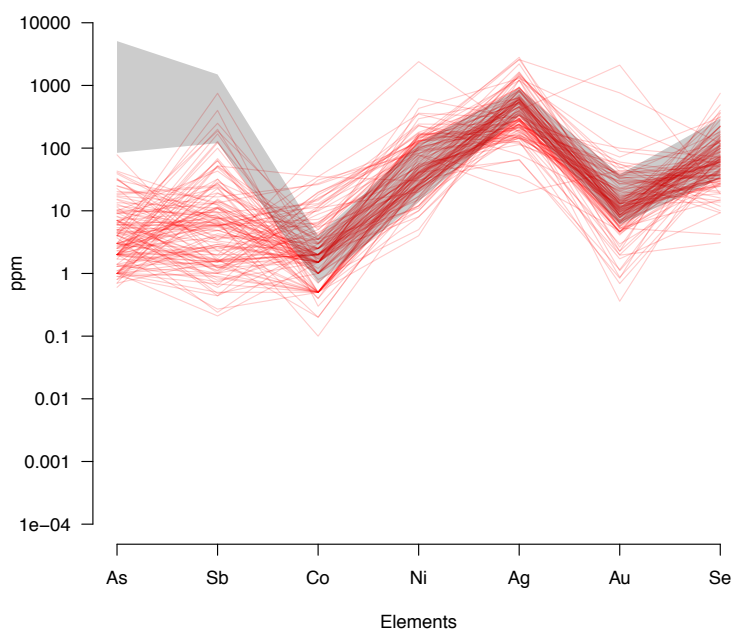

**Figure S3. Trace element signatures of 129 copper artefacts belonging to cluster 4 (lines) plotted against the trace element signature of Ai Bunar (grey field), a copper mine in central Bulgaria.**

**Sites Network – community structure analyses of archaeological sites**. In this step, the archaeological sites represented nodes, and links between them were based on sharing the same chemical cluster for pairs of copper artefacts found in those sites. This relationship was established under the assumption that two artefacts belonging to the same chemical cluster could have ended up from the places of exploitation or production in two different sites through either direct or indirect contact (i.e. various types of intermediaries); we encompass both options under the term 'supply network'. Thus, the link between the sites in out network practically works in the following way: artefact A and artefact B from two different sites belong to (chemical) cluster 1, and therefore these two sites (nodes) have a link placed between them. If these two sites contain more artefacts from the same cluster, the weight of the link is larger. For example: if site $i$ contains artefacts from clusters [0,1,1,1,1,2,2,2,3] and site $j$ has artefact from clusters [0,1,1,2,2,8,9], then the weight of the link is 5 (one for each artefact of the common type). We analysed the final network with Louvain algorithm [35] and gained only one level with three distinctive community structures.

When randomizing the network, we cut each link and randomly reconnected it to a different node while saving only the information of the degree of each node for this type of network. We took into consideration, for instance, that the link with weight 5 is actually 5 links. We repeat the randomization procedure 1000 times. The modularity of the original network (Sites Network) was 0.276, which is 57 standard deviations larger from the mean of the modularities of the randomized network (0.078 ± 0.004) (Figure 3b). Geographical coordinates of archaeological sites/nodes (Table S1) were used solely for illustrative purposes in this paper. Visualisation of Sites Network (Figures 4-7) was produced in Python from scratch, using Matlibplot package and the background map with kind permission of Prof. M. Milinkovic (University of Belgrade, Serbia). The Sites Network is the final outcome of our network design, and the only one whose modularity we discuss in the article.

Since some of the observed sites (nodes) were active throughout multiple time-periods, and we wanted to observe their position in each of them, we regarded the same site in a different period as a separate node (site-period), and added the chronological span to the site name for easier navigation through results (see Table S2). Most importantly, apart from chemical cluster number we did not use any archaeologically relevant information in our network. In total, we have 79 sites and 93 site-periods. The sites (nodes) that appear in more than one period are listed below (Figure S4):

- Ai Bunar   4600-4450 BC, 4100-3700 BC

- Belovode 5500-5000 BC, 5000-4600 BC, 4600-4450 BC

- Durankulak 5000-4600 BC, 4600-4450 BC, 4450-4100 BC, 3700-3200 BC

- Goljamo Delcevo 4600-4450 BC, 4450-4100 BC

- Gomolava 5000-4600 BC, 4600-4450 BC

- Hotnica 4450-4100 BC, 3700-3200 BC

- Pločnik 5500-5000 BC, 5000-4600 BC, 4600-4450 BC

- Smjadovo 4450-4100 BC, 3700-3200 BC

- Tell Ruse 4600-4450 BC, 4450-4100 BC

- Zlotska pecina 4100-3700 BC, 3700-3200 BC

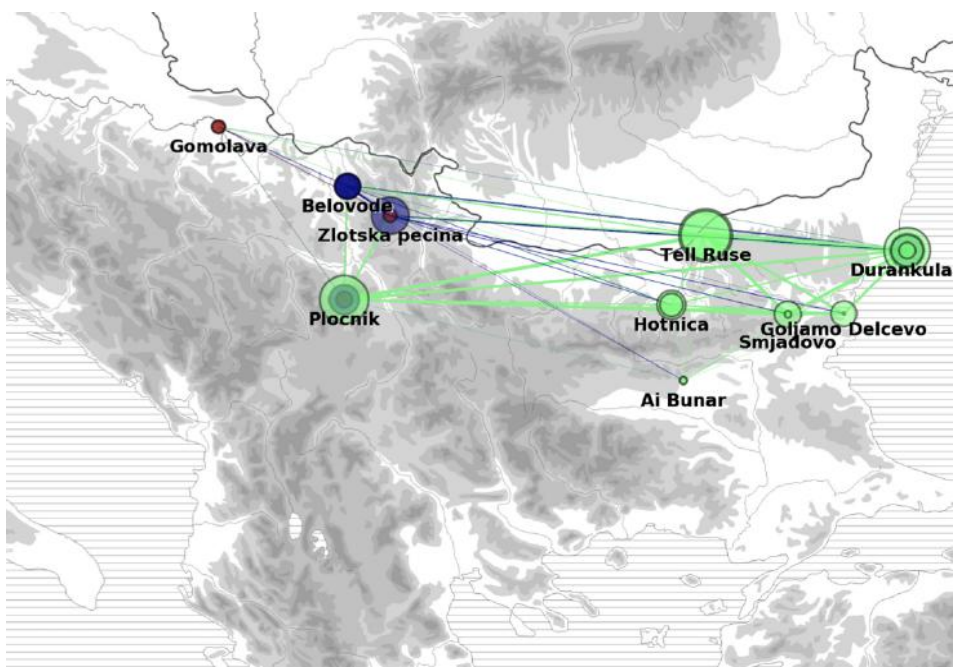

**Figure S4. The sites (nodes) that exist throughout multiple time periods. Pločnik, Zlotska pecina and Goljamo Delcevo change the module over the observed time frame (c. 6200 to c. 3200 BC).**

**The importance of archaeological sites (nodes).** We initially tested the importance of our nodes with three different node centrality measurements: degree centrality (based on number of links each node includes), PageRank [47] and betweenness centrality [48]. All three yielded meaningful results for determining the importance of the specific archaeological sites. The degree centrality of the node (in this case weighted degree or strength) tells us with how many other sites the observed site had some kind of communication. The PageRank takes into account how important the observed sites are. However, given that our network is not directed, these two properties appear significantly correlated (see Figure S5), and hence both presented similar results for our study. On the other hand, the betweenness centrality is defined as a number of shortest paths that go through an observed node. In order to calculate it we defined the weights as $1/w$ or $1/w^2$, where $w$ is the weight in the original network; this procedure ensured that if there were more connections between the sites, it was easier to travel between them. Once we compared the betweenness centrality and the PageRank we observed that barring the large difference for nodes of smaller PageRank values, the more important nodes were still more important by both measures (see Figure S6). Also, the betweenness centrality measure is not very robust and by removing only one artefact from the original input, the values change substantially, although again the more important nodes still come out the same. To conclude, using any of the importance measure yielded very similar results, which is why we give all three in Table S2. For the purpose of illustration in our maps (size of the nodes) we opted for PageRank; these are, again, not robust, which is why we use them only for visualisation.

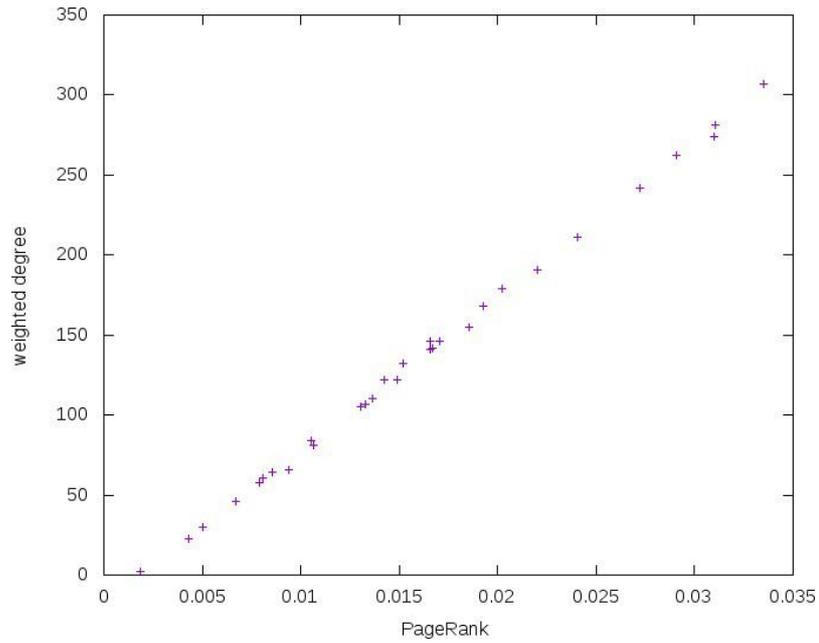

**Figure S5. PageRank vs. weighted degree (strength). The two measures are strongly correlated, as expected in undirected networks, which makes both useful for measuring the importance of the site (node).**

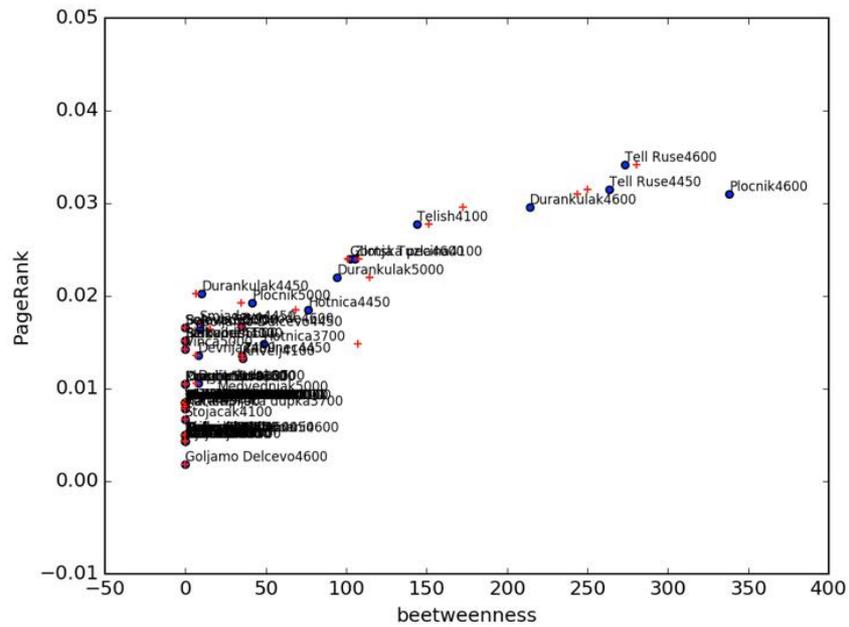

**Figure S6. PageRank vs. betweenness centrality. Please note that except for the values with small PageRank, the two measures are correlated, which makes both suitable for measuring importance of the site.**